\documentclass[pop,fleqn]{w-art}
\usepackage{times}
\usepackage{w-thm}
\usepackage{amsmath}
\usepackage{amsfonts} 
\usepackage{amssymb} 

\theoremstyle{plain}

\theoremstyle{definition}

\usepackage{graphicx}
\newcommand{\eq}[1]{Eq.~(\ref{#1})}
\def\ee{\mathrm{e}}

\def\one{{\rm 1\kern -.9mm l}}

\def\ii{{\mbox{i}}}

\newcommand{\lvev}{\Big\langle\hskip -5pt\Big\langle}
\newcommand{\rvev}{\Big\rangle\hskip -5pt\Big\rangle}

\begin{document}
\DOIsuffix{theDOIsuffix}
\Volume{51}
\Issue{1}
\Month{01}
\Year{2003}
\pagespan{1}{}



\title[Instanton Calculus With R-R Background]{Instanton Calculus With
R-R Background And Topological Strings}


\author[M. Bill\'o]{Marco Bill\'o\inst{1}%
  \footnote{Corresponding author\quad E-mail:~\textsf{billo@to.infn.it},
            Phone: +0039\,011\,670\,7213,
            Fax: +0039\,011\,670\,7213}}
\address[\inst{1}]{Dip. di Fisica Teorica, Universit\`a di Torino
and I.N.FN., sez. di Torino\\ 
Via Pietro Giuria 1, 10125 Torino (ITALY)}
\begin{abstract}
  The non-perturbative low energy effective action of N=2 SYM is studied within a
  microscopic string realization via D3/D-instanton systems. The localization
  deformation of instanton moduli space which has allowed the exact computation of
  multi-instanton contributions
  is, in this setting, due to a RR graviphoton background.
  The relation of deformed instanton contributions to topological string
  amplitudes on CY, argued by Nekrasov, appears quite natural in this framework.
  Based on \cite{nuovo}.
\end{abstract}
\maketitle                   





\section{Introduction}
Recently, D-instanton-induced interactions in the low energy effective actions
are receiving quite some attention. In this contribution,
we study {D-instanton} induced  couplings of the  {chiral} and the 
{Weyl} multiplet in $\mathcal{N}=2$ SYM low energy effective theory. 
In this framework, we obtain a natural interpretation of a
conjecture by Nekrasov regarding the {$\mathcal{N}=2$ 
multi-instanton calculus} and its relation to 
{topological string} amplitudes on {CY}'s 

The semiclassical limit of the Seiberg-Witten prepotential displays 
{1-loop} plus {instanton} contributions:
\begin{equation*}
\label{Fexp}
{\mathcal{F}}({a}) = 
{\frac{\ii}{2\pi} {a^2} \log \frac{{a^2}}{\Lambda^2}} 
+
{\sum_{k=1}^\infty \mathcal{F}^{(k)}({a})}~.
\end{equation*}
After the Seiberg-Witten solution was put forward, it became very important
to check it against the computation of the multi-instanton contributions
{$\mathcal{F}^{(k)}({a})$} within the {``microscopic''} 
description of the non-abelian gauge theory.
The calculation of these coefficients has 
only recently been fully accomplished \cite{Nekrasov:2002qd,Flume:2002az,Losev:2003py}.
Introducing a suitable {deformation} of the {ADHM measure} on the 
super-instanton {moduli spaces}
which exploits the 4d chiral rotations symmetry of ADHM constraints, the resulting
partition function 
can be computed using {localization} techniques and the topological twist of 
its supersymmetries. One has
\begin{equation*}
Z({a},{\varepsilon}) = \sum_k Z^{(k)}({a},{\varepsilon})
=\sum_k \int d{\widehat{\mathcal{M}}_{(k)}} 
\ee^{-\mathcal{S}_{\mbox{\tiny mod}}({a},{\varepsilon};{\mathcal{M}_{(k)}})}
= \exp\left(\frac{\mathcal{F}_{\mbox{\tiny n.p.}}
({a};{\varepsilon})}{{\varepsilon^2}}\right)
\end{equation*}
and 
$\lim_{{\varepsilon \to 0}} \mathcal{F}_{\mbox{\tiny n.p.}}({a};{\varepsilon}) =
\mathcal{F}_{\mbox{\tiny n.p.}}({a})$ 
yields the non-perturbative part of the SW prepotential.
What about higher orders in the deformation parameter {$\varepsilon$}?

Nekrasov proposed \cite{Nekrasov:2002qd,Losev:2003py,Nekrasov:2005wp} that terms of order 
${\varepsilon}^{{2h}}$ correspond to {gravitational}
$F$-terms in the {$\mathcal{N}=2$ eff. action} involving {metric} and 
{graviphoton} curvatures: 
\begin{equation*}
\int d^4x ({R^+})^{{2}}  ({\mathcal{F}^{+}})^{{2h-2}}~.
\label{fh}
\end{equation*}
When the effective {$\mathcal{N}=2$ theory} is obtained from 
{type II strings} on a ``local'' 
{CY} manifold {$\mathfrak{M}$} via geometrical engineering \cite{Kachru:1995fv,Katz:1996fh}, 
such terms arise from world-sheets of genus {$h$} and are computed by the {topological string}
\cite{Bershadsky:1993cx,Antoniadis:1993ze}.  
For the {local CY} describing the $\mbox{SU}(2)$ theory
the proposal has been tested in \cite{Klemm:2002pa}.

We will discuss how to
reproduce the semiclassical {instanton expansion} of the 
{low energy effective action} for the $\mathcal{N}=2$ SYM theory
via its microscopic string realization via  (fractional) 
{D3}/{D(-1)} branes. We will then show 
that the inclusion of the {graviphoton} of the $\mathcal{N}=2$ 
bulk sugra, which comes from the {RR} sector, 
produces in the effective action the {gravitational F-terms} which are computed by the 
{topological string} on {local CY}; at the same time, it
leads exactly to the {localization deformation} on the instanton {moduli space}
which allows to perform the integration.


\section{Stringy instanton calculus for $\mathcal{N}=2$ SYM}

Consider pure $\mbox{SU}(N)$ Yang-Mills in 4 dimensions 
with {$\mathcal{N}=2$} susy.
It is realized by the massless d.o.f. of {open strings} attached to
{fractional D3-branes} in the {orbifold} background
${\mathbb{R}^4} \times \mathbb{R}^2 \times 
{\mathbb{R}^4/\mathbb{Z}_2}$
These d.o.f. can be can be arranged in a ${\cal N}=2$ chiral superfield
in the adjoint of $\mathrm{U}(N)$:
\begin{equation}
\label{Pm1}
\Phi(x,\theta) = {\phi(x)} + \theta{\Lambda(x)} +\frac
12\,\theta\sigma^{\mu\nu} \theta
\,{F_{\mu\nu}^+(x)} + \cdots~.
\end{equation}
String amplitudes for the string vertices corresponding to the 
SYM fields of \eq{Pm1}
on {discs} attached to the {D3 branes}
give rise, in the limit
{$\alpha'\to 0$} with {gauge quantities fixed},
to the tree level (microscopic) {$\mathcal{N}=2$} action for $\mathrm{U}(N)$ SYM,
where $N$ is the number of fractional D3-branes.
We are interested in the  l.e.e.a. on the {Coulomb branch} 
parametrized by the {v.e.v.'s} 
${\langle \Phi_{uv} \rangle} = {a_u}\,\delta_{uv}$
of the adjoint chiral superfields
breaking $\mbox{SU}(N) \to  \mbox{U}(1)^{N-1}$;
we focus for simplicity on $\mbox{SU}(2)$.
Up to two-derivatives, $\mathcal{N}=2$ susy forces
the effective action for the chiral multiplet {$\Phi$} in the Cartan direction
to be of the form
\begin{equation*}
\label{Seff}
S_{\mbox{\tiny eff}}[{\Phi}] = 
\int d^4x \,d^4\theta\, {\mathcal{F}}({\Phi}) + \mathrm{c.c}~,
\end{equation*}
where $\mathcal{F}$ is called the prepotential of the theory.
We want now to discuss how the {instanton corrections} 
to the {prepotential} arise
in our {string set-up}.

Since the {topological density} of an instantonic configuration
corresponds to a localized source for the RR scalar ${C_0}$ in the WZ part of
the D3-brane action, 
{instanton-charge $k$} solutions of 3+1 dims. $\mbox{SU}(N)$ 
gauge theories correspond to  {$k$ D-instantons} inside $N$
D3-branes
\cite{Witten:1995im,Douglas} 

Open strings ending on a {D(-1)} carry {no momentum}:
the polarizations of their physical vertices are 
{moduli} (rather than {fields}), and they correspond to the
parameters of the instanton, see Table \ref{table:moduli}; we will denote them as 
{$\mathcal{M}_{(k)}$}~.
\begin{vchtable}
\vchcaption{The spectrum of moduli for the fractional D(-1)/D3 system}
{\footnotesize
\begin{tabular}{@{}c|cccc}
 & ADHM  & Meaning & Vertex & Chan-Paton \\
\hline  
{-1/-1} (NS) & ${a'_\mu}$ & \emph{centers} &
$\psi^{\mu}(z)\ee^{-\varphi(z)}$ & adj. ${\mbox{U}(k)}$ \\
   & ${\chi}$ & \emph{aux.} &
${\overline\Psi}(z)\ee^{-\varphi(z)}$ & 
   $\vdots$ \\
(aux. vert.) &  {$D_c$} & \emph{Lagrange mult.} &
$\bar\eta_{\mu\nu}^c\psi^\nu(z)\psi^\mu(z)$ & $\vdots$ \\
\phantom{-1/-1} (R) & ${{M}^{\alpha A}}$ &  \emph{partners} & 
$S_{\alpha}(z) S_{A}(z)\ee^{-\frac{1}{2}\varphi(z)}$ & $\vdots$ \\
   & ${\lambda_{\dot\alpha A}}$ & \emph{Lagrange mult.} &
$S^{\dot\alpha}(z)S^{A}(z)\ee^{-\frac{1}{2}\varphi(z)}$ & $\vdots$\\
\hline
{-1/3} (NS) &  ${{w}_{\dot\alpha}}$ & \emph{sizes} &
${\Delta}(z) S^{\dot\alpha}(z)\,\ee^{-\varphi(z)}$ &
${k}\times {\overline N}$\\
  & ${{\bar w}_{\dot\alpha}}$ & $\vdots$ &
${\overline\Delta}(z) S^{\dot\alpha}(z)\ee^{-\varphi(z)}$ & $\vdots$\\
\phantom{-1/3} (R) & ${{\mu}^A}$ & \emph{partners}  &
${\Delta}(z) S_{A}(z)\ee^{-{\frac12}\varphi(z)}$ & $\vdots$\\
  & ${{\bar \mu}^A}$ & $\vdots$ & 
${\overline\Delta}(z) S_{A}(z)\ee^{-{\frac12}\varphi(z)}$ & $\vdots$\\
\hline 
\end{tabular}
}
\label{table:moduli}
\end{vchtable}
\begin{vchfigure}[t]
\begin{picture}(0,0)%
\includegraphics{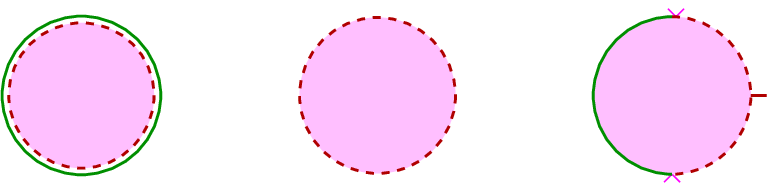}%
\end{picture}%
\setlength{\unitlength}{1657sp}%
\begingroup\makeatletter\ifx\SetFigFont\undefined%
\gdef\SetFigFont#1#2#3#4#5{%
  \reset@font\fontsize{#1}{#2pt}%
  \fontfamily{#3}\fontseries{#4}\fontshape{#5}%
  \selectfont}%
\fi\endgroup%
\begin{picture}(9749,3094)(197,-2459)
\put(7786,479){\makebox(0,0)[lb]{\smash{{\SetFigFont{5}{6.0}{\rmdefault}{\mddefault}{\updefault}{\small $\bar w$}}}}}
\put(9946,-691){\makebox(0,0)[lb]{\smash{{\SetFigFont{5}{6.0}{\rmdefault}{\mddefault}{\updefault}$+ \ldots $}}}}
\put(2566,-646){\makebox(0,0)[lb]{\smash{{\SetFigFont{5}{6.0}{\rmdefault}{\mddefault}{\updefault}$\equiv$}}}}
\put(5986,-646){\makebox(0,0)[lb]{\smash{{\SetFigFont{5}{6.0}{\rmdefault}{\mddefault}{\updefault}$+$}}}}
\put(8821,-1006){\makebox(0,0)[lb]{\smash{{\SetFigFont{5}{6.0}{\rmdefault}{\mddefault}{\updefault}{\small $\lambda$}}}}}
\put(3871,-2401){\makebox(0,0)[lb]{\smash{{\SetFigFont{5}{6.0}{\rmdefault}{\mddefault}{\updefault}{\small $-\frac{8\pi^2\,k}{g^2}$}}}}}
\put(8011,-2401){\makebox(0,0)[lb]{\smash{{\SetFigFont{5}{6.0}{\rmdefault}{\mddefault}{\updefault}{\small $\mathcal{S}_{\mbox{\tiny mod}}$}}}}}
\put(7741,-1861){\makebox(0,0)[lb]{\smash{{\SetFigFont{5}{6.0}{\rmdefault}{\mddefault}{\updefault}{\small $\mu$}}}}}
\put(2251,-2401){\makebox(0,0)[lb]{\smash{{\SetFigFont{5}{6.0}{\rmdefault}{\mddefault}{\updefault}{$\stackrel{\scriptsize\alpha'\to 0}{\simeq}$}}}}}
\end{picture}%
 \vchcaption{D-instanton discs contributing to the partition function.}
\label{fig:dm}
\end{vchfigure}

Let us consider first disc diagrams involving only {moduli} 
and {no D3/D3 state}; these are ``vacuum'' contributions from the 
{D3} point of view; see Fig. \ref{fig:dm} 
The combinatorics of boundaries \cite{Polchinski:fq}
is such that these D-instanton diagrams {exponentiate}. Moreover,
the {moduli} must be {integrated over} to produce the partition function
\begin{equation*}
\label{Zk}
Z^{(k)} = 
\int d{\mathcal{M}_{(k)}} 
\ee^{-\frac{8\pi^2\,k}{{g^2}} - \mathcal{S}_{\mbox{\tiny mod}}}~.
\end{equation*}

From disc diagrams with insertion of {moduli} vertices, 
in the field theory limit we extract the {ADHM} moduli action (at fixed $k$):
$
{\mathcal{S}_{\mbox{\tiny mod}}}=
{\mathcal{S}_{\mbox{\tiny bos}}^{(k)}}
+ {\mathcal{S}_{\mbox{\tiny fer}}^{(k)}}
+ \mathcal{S}_{\mbox{\tiny c}}^{(k)}
$,
with 
\begin{eqnarray}
\label{smod}
{\mathcal{S}_{\mbox{\tiny bos}}^{(k)}} & = & 
{\mbox{tr}_k\Big\{ -2\,[\chi^{\dagger},a'_\mu][\chi,{a'}^\mu] +
\chi^{\dagger}{\bar w}_{\dot\alpha} w^{\dot\alpha}\chi
+ \chi{\bar w}_{\dot\alpha} w^{\dot\alpha} \chi^{\dagger}\Big\}}~,
\nonumber\\
{\mathcal{S}_{\mbox{\tiny fer}}^{(k)}} & = & 
{\mbox{tr}_k\Big\{\ii\,
\frac{\sqrt 2}{2}\,{\bar \mu}^A \epsilon_{AB} \mu^B\chi^{\dagger}
-\ii\,
\frac{\sqrt 2}{4}\,M^{\alpha A}\epsilon_{AB}[\chi^{\dagger},M_{\alpha}^{B}]
\Big\}}~,
\nonumber\\
\mathcal{S}_{\mbox{\tiny c}}^{(k)} & = & \mbox{tr}_k\Big\{\!-\ii D_c
\big({{W}^c +\ii
\bar\eta_{\mu\nu}^c \big[{a'}^\mu,{a'}^\nu\big]}\big) \notag
\ii {\lambda}^{\dot\alpha}_{\,A}\big(
{\bar{\mu}^A{w}_{\dot\alpha}+
\bar{w}_{\dot\alpha}{\mu}^A  +
\big[a'_{\alpha\dot\alpha},{M'}^{\alpha A}\big]\big)}\!
\Big\}~.
\end{eqnarray}
$\mathcal{S}_{\mbox{\tiny c}}^{(k)}$ displays the {bosonic} and {fermionic} 
{ADHM constraints}.

Consider now correlators of {D3/D3} fields, e.g of the scalar ${\phi}$
in the Cartan direction, in presence of {$k$ D-instantons}. It turns 
out \cite{Green:1997tv,Green:2000ke,Billo:2002hm}
that
the dominant contribution to 
$\langle{\phi_1}\ldots {\phi_n}\rangle$ is from 
{$n$ one-point} amplitudes on 
discs with moduli insertions. 
The result can therefore be encoded in extra moduli-dependent vertices for {$\phi$'s}, i.e. in 
{extra terms} in the moduli action containing such {one-point} functions: 
\begin{equation*}
 \mathcal{S}_{\mbox{\tiny mod}}({\varphi};{\mathcal{M}}) = 
{\phi}({x}) J_\phi({\widehat{\mathcal{M}}}) + 
\mathcal{S}_{\mbox{\tiny mod}}({\widehat{\mathcal{M}}})~,
\end{equation*}
where {$x$} is the instanton center and 
${\phi}({x}) J_\phi({\widehat{\mathcal{M}}})$ 
is given by the disc diagrams with boundary (partly) on the D(-1)'s describing 
the emission of a $\phi$.
To determine  
$\mathcal{S}_{\mbox{\tiny mod}}({\phi};{\mathcal{M}})$ 
we systematically compute mixed discs with a scalar
{$\phi$} emitted from the D3 boundary, such as the one of Fig. \ref{fig:xphiw}

Other non-zero diagrams, related 
by the Ward identities of the susies broken by the D(-1),
couple the other components of the {gauge supermultiplet} to the {moduli}. 
The superfield-dependent moduli action
$\mathcal{S}_{\mbox{\tiny mod}}({\Phi};{\mathcal{M}})$
is thus obtained by simply letting 
${\phi(x)} \rightarrow {\Phi}({x},\theta)$.

\section{Inclusion of a graviphoton background}

In the stringy setup, is quite natural to consider also the effect
of {D-instantons} on
correlators of fields from the {closed
string} sector.
The effect can be encoded in a field-dependent moduli action
determined from one-point functions of {closed string vertices} 
on {instanton discs} with {moduli} insertions.
Our aim is to study interactions in the low energy 
{$\mathcal{N}=2$} effective action involving the {graviphoton}.

\begin{figure}
\begin{minipage}[b]{0.45\textwidth}
\begin{picture}(0,0)%
\includegraphics{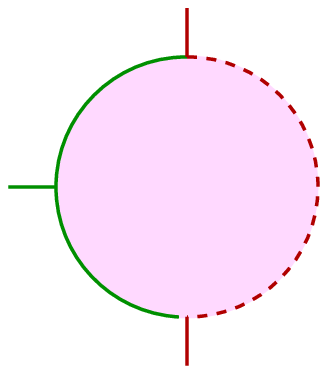}%
\end{picture}%
\setlength{\unitlength}{1816sp}%
\begingroup\makeatletter\ifx\SetFigFont\undefined%
\gdef\SetFigFont#1#2#3#4#5{%
  \reset@font\fontsize{#1}{#2pt}%
  \fontfamily{#3}\fontseries{#4}\fontshape{#5}%
  \selectfont}%
\fi\endgroup%
\begin{picture}(3033,3813)(376,-3208)
\put(2326,389){\makebox(0,0)[lb]{\smash{{\SetFigFont{8}{9.6}{\familydefault}{\mddefault}{\updefault}${\bar X^\dagger}$}}}}
\put(1651,-3136){\makebox(0,0)[lb]{\smash{{\SetFigFont{8}{9.6}{\familydefault}{\mddefault}{\updefault}$w$}}}}
\put(376,-1636){\makebox(0,0)[lb]{\smash{{\SetFigFont{8}{9.6}{\familydefault}{\mddefault}{\updefault}${\phi}$}}}}
\end{picture}%
\caption{Emission of a D3 field from a mixed disc.}
\label{fig:xphiw}
\end{minipage}
\hfil
\begin{minipage}[b]{0.50\textwidth}
\begin{picture}(0,0)%
\includegraphics{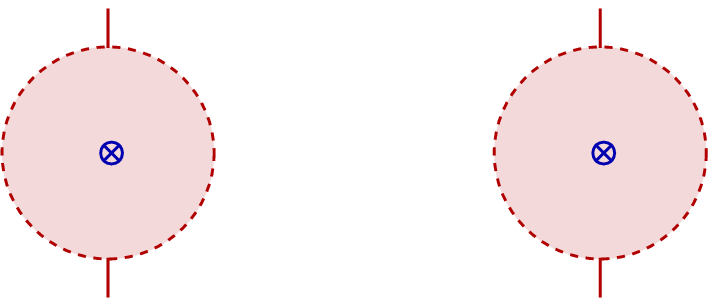}%
\end{picture}%
\setlength{\unitlength}{1658sp}%
\begingroup\makeatletter\ifx\SetFigFont\undefined%
\gdef\SetFigFont#1#2#3#4#5{%
  \reset@font\fontsize{#1}{#2pt}%
  \fontfamily{#3}\fontseries{#4}\fontshape{#5}%
  \selectfont}%
\fi\endgroup%
\begin{picture}(8095,3822)(116,-3067)
\put(826,-2986){\makebox(0,0)[lb]{\smash{{\SetFigFont{8}{9.6}{\familydefault}{\mddefault}{\updefault}{$a'$}}}}}
\put(1501,539){\makebox(0,0)[lb]{\smash{{\SetFigFont{8}{9.6}{\familydefault}{\mddefault}{\updefault}{$Y^\dagger$}}}}}
\put(6451,-2986){\makebox(0,0)[lb]{\smash{{\SetFigFont{8}{9.6}{\familydefault}{\mddefault}{\updefault}{$M$}}}}}
\put(7126,539){\makebox(0,0)[lb]{\smash{{\SetFigFont{8}{9.6}{\familydefault}{\mddefault}{\updefault}{$M$}}}}}
\put(1051,-811){\makebox(0,0)[lb]{\smash{{\SetFigFont{8}{9.6}{\familydefault}{\mddefault}{\updefault}{$\mathcal{F},\bar{\mathcal{F}}$}}}}}
\put(6751,-811){\makebox(0,0)[lb]{\smash{{\SetFigFont{8}{9.6}{\familydefault}{\mddefault}{\updefault}{$\bar{\mathcal{F}}$}}}}}
\end{picture}%
\caption{Graviphoton emission from an instanton disc.}
\label{fig:disk_grav}
\end{minipage}
\end{figure}

The field content of {$\mathcal{N}=2$ sugra}, namely the metric 
$h_{\mu\nu}$, the gravitini $\psi_\mu^{\alpha A}$ and the graviphoton $C_\mu$
can be organized in a {chiral Weyl multiplet}:
\begin{equation*}
{W^+_{\mu\nu}(x,\theta)}= {\mathcal{F}_{\mu\nu}^+(x)} +
\theta {\chi_{\mu\nu}^+(x)}+\frac{1}{2}
\,\theta\sigma^{\lambda\rho} \theta\,{R^+_{\mu\nu\lambda\rho}(x)}
+\cdots
\label{weyl}
\end{equation*}
({$\chi_{\mu\nu}^{~~\alpha A}$} is the gravitino field strength).
These fields arise from massless vertices of {type IIB strings} on 
${\mathbb{R}^{4}}\times \mathbb{C}\times {\mathbb{C}^2/\mathbb{Z}_2}$.
In particular, the graviphoton vertex
is given by%
\footnote{A {different} RR field, with a {similar} structure, will be
useful ($\hat A, \hat B=3,4$ $\leftrightarrow$  are {odd} ``internal'' 
spin fields):
\begin{equation*}
{V_{\bar{\mathcal{F}}}}(z,\bar z) \! = \!  
\frac{1}{4\pi}
{\bar{\mathcal{F}}^{\alpha\beta \hat A\hat B}}(p)
\!\times\!
\Big[S_\alpha(z)S_{\hat A}(z)\ee^{-\frac{1}{2}\varphi(z)}
{S}_\beta(\bar z){S}_{\hat B}(\bar z)\ee^{-\frac{1}{2}{\varphi}(\bar
z)}\Big]\ee^{\ii p\cdot X(z,\bar z)}~.
\label{Vbarf}
\end{equation*}
}
\begin{equation*}
{V_{\mathcal{F}}}(z,\bar z)  \!=\!  \frac{1}{4\pi}{\mathcal{F}^{\alpha\beta AB}}(p)
\!\times\!
\Big[S_\alpha(z)S_A(z)\ee^{-\frac{1}{2}\varphi(z)}
{S}_\beta(\bar z){S}_B(\bar z)\ee^{-\frac{1}{2}{\varphi}(\bar
z)}\Big]\ee^{\ii p\cdot X(z,\bar z)}~,
\end{equation*}
left-right movers identification on discs being taken into account.
The bi-spinor graviphoton polarization is given by
$
{\mathcal{F}^{(\alpha\beta) [AB]}} = \frac{\sqrt 2}{4}\,
{\mathcal{F}_{\mu\nu}^+}\big(\sigma^{\mu\nu})^{\alpha\beta}\,\epsilon^{AB}
$
and corresponds to a RR 3-form $\mathcal{F}_{\mu\nu z}$ with one ``leg'' in the
$\mathbb{C}$ internal direction. 
To determine the contribution of the graviphoton to the field-dependent
moduli action we have to consider disc amplitudes with open string 
{moduli vertices} on the boundary and closed string {graviphoton 
vertices} in the interior which survive in the field theory limit $\alpha'\to 0$. 
Other diagrams, connected by susy, have the effect of promoting the dependence
of the moduli action to the full Weyl multiplet:
${\mathcal{F}^+_{\mu\nu}} \to {W^+_{\mu\nu}}(x,\theta)$.
It turns out that very few diagrams (depicted in Fig. \ref{fig:disk_grav}) 
contribute.

These diagrams are easily evaluated; for instance, one has 
\begin{equation*}
\lvev {V_{M} V_{M}}{V_{\bar{\mathcal F}}} \rvev
=\frac{1}{4\sqrt 2}\mbox{tr}_k\Big\{{M^{\alpha A}M^{\beta B}}
{\bar{\mathcal{F}}^+_{\mu\nu}}
\Big\}(\sigma^{\mu\nu})_{\alpha\beta}\epsilon_{AB}~.
\label{mmbare}
\end{equation*}
Integrating over the {moduli} the interactions described by 
the field-dependent moduli action\\ 
$\mathcal{S}_{\mbox{\tiny mod}}({\Phi},{W^+};{\mathcal{M}_{}(k)})$
one gets the {effective action}, and hence the prepotential, for the long-range multiplets 
{$\Phi$} and {$W^+$} in the  {instanton number $k$ } sector:
\begin{equation*}
\label{effacW}
 S_{\mbox{\tiny eff}}^{(k)}[{\Phi},{W^+}] \equiv
\int d^4x \, d^4\theta\, \mathcal{F}^{(k)}({\Phi},{W^+}) =
\int d^4x \, d^4\theta\,
d{\widehat{\mathcal{M}}_{(k)}}\, \ee^{-\frac{8\pi k}{g^2} -
\mathcal{S}_{\mbox{\tiny mod}}({\Phi},{W^+};{\mathcal{M}_{}(k)})}~.
\end{equation*}
{$\Phi(x,\theta)$} and {$W_{\mu\nu}^+(x,\theta)$} are constant 
w.r.t. the integration variables ${\widehat{\mathcal{M}}_{(k)}}$.
We can compute $\mathcal{F}^{(k)}({a};{f})$ giving them
{constant values}: ${\Phi}(x,\theta) \to {a}$,
${W^+_{\mu\nu}}(x,\theta) \to {f_{\mu\nu}}$,
getting a ``deformed'' moduli action
\begin{eqnarray*}
\label{defmodac}
&&\hskip -0.2cm
\mathcal{S_{\mbox{\tiny mod}}}({a,\bar a};{f,\bar f}) = 
{S^{(k)}_{\mbox{\tiny c}}}
- \mbox{tr}_k\Big\{
\big([\chi^{\dagger},a'_{\alpha\dot\beta}]+2{\bar f_c}(\tau^c a')_{\alpha\dot\beta}\big)
\big([\chi,{a'}^{\dot\beta\alpha}]+2{f_{c}}(a'\tau^c)^{\dot\beta \alpha}\big)
\nonumber\\
&&\hskip -0.2cm
-\big(\chi^{\dagger}{\bar w}_{\dot\alpha}-{\bar w}_{\dot\alpha}\,{\bar a}\big)
\big( w^{\dot\alpha}\chi- {a}\,w^{\dot\alpha}\big)
-\big(\chi{\bar w}_{\dot\alpha} -{\bar w}_{\dot\alpha}\,{a}\big)
\big(w^{\dot\alpha}\chi^{\dagger}
-{\bar a}\,w^{\dot\alpha}\big) \Big\}
\nonumber\\
&&\hskip -0.2cm
+\ii\, \frac{\sqrt 2}{2}\,
\mbox{tr}_k\Big\{{\bar \mu}^A \epsilon_{AB} \big( \mu^B\chi^{\dagger}
-{\bar a}\,\mu^B\big)
-\frac{1}{2}\,M^{\alpha A}\epsilon_{AB}\big([\chi^{\dagger},M_{\alpha}^{B}]
+2\,{\bar f_c}\, (\tau^c)_{\alpha\beta}M^{\beta B}\big)
\Big\}~.
\end{eqnarray*}
Notice that the constraint part of the action, ${S^{(k)}_{\mbox{\tiny c}}}$, 
is not modified.

In the action 
${\mathcal{S}_{\mbox{\tiny mod}}}({a,\bar a};{f,\bar f})$
the v.e.v.'s ${a},{f}$ and ${\bar a},{\bar f}$ are not on the same footing.
Indeed, we have 
$\mathcal{S}_{\mbox{\tiny mod}}({a,\bar a};{f,\bar f}) = {Q}\,\Xi$
where {$Q$} is the scalar component
${Q} \equiv 
\frac{1}{2}\,\epsilon_{\dot\alpha\dot\beta}\,Q^{\dot\alpha\dot\beta}$
of the twisted supercharges
$Q^{\dot\alpha B}\stackrel{\mbox{\tiny top. twist}}{\longrightarrow}
Q^{\dot\alpha\dot\beta}$. The parameters {$\bar a$},{$\bar f_c$} appear {only}
in the gauge fermion $\Xi$. The instanton partition function
\begin{equation*}
Z^{(k)}\equiv \int d\mathcal{M}_{(k)}\,
\ee^{-\mathcal{S}_{\mbox{\tiny mod}}({a,\bar a};{f,\bar f})}
\label{partfunct}
\end{equation*}
is therefore {independent} of {$\bar a$},{$\bar f_c$}: its variation w.r.t
these parameters is {$Q$-exact}.

The moduli action obtained inserting the {graviphoton} background
coincides {exactly} with the {``deformed''} action
\cite{Nekrasov:2002qd,Flume:2002az,Losev:2003py,Nekrasov:2005wp}
 considered in the literature
to localize the moduli space integration if we set
\begin{equation}
\label{floc}
{f_c}=\frac{{\varepsilon}}{2}\,\delta_{3c}~,~~
{\bar f_c} = \frac{{\bar\varepsilon}}{2}\,\delta_{3c}~,~~
\end{equation}
and moreover (referring to the notations in the above ref.s)
${\varepsilon}={\bar \varepsilon}$~,
${\varepsilon} = {\epsilon_1}=-{\epsilon_2}$.
From the explicit form of $\mathcal{S_{\mbox{\tiny mod}}}({a,0};{f,0})$
it follows that the prepotential $\mathcal{F}^{(k)}({a};{f})$ is invariant under
${a}, {f_{\mu\nu}} \to {-a}, -{f_{\mu\nu}}$.
It must moreover be regular for ${f}\to 0$, to 
reproduce the {instanton \# $k$} contribution to the SW 
prepotential. Since odd powers of ${a}{f}_{\mu\nu}$ cannot appear, 
one has altogether, reinstating the superfields, 
\begin{equation*}
\label{Fkpw}
\mathcal{F}^{(k)}({\Phi}, {W^+}) =  \sum_{h=0}^\infty {c_{k,h}} \,
{\Phi^2}
\left(\frac{\Lambda}{{\Phi}}\right)^{4k}\!\left(\frac{{W^+}}{{\Phi}}\right)^{2h}~.
\end{equation*}
 Summing over the instanton sectors we obtain
\begin{equation}
\label{Fnp}
\mathcal{F}_{\mbox{\tiny n.p.}}({\Phi},{W^+}) = 
\sum_{k=1}^\infty \mathcal{F}^{(k)}({\Phi},{W^+})
= \sum_{{h}=0}^\infty C_{{h}}(\Lambda,{\Phi}) {(W^+)^{2h}}
~,~~~
C_{{h}}(\Lambda,{\Phi}) = 
\sum_{k=1}^\infty {c_{k,h}} \,\frac{\Lambda^{4k}}{{\Phi}^{4k+2{h}-2}}~.
\end{equation}
This yields many different terms in the effective action,
connected by susy. In particular, saturating the $\theta$ integration with
four $\theta$'s all from ${W^+}$ we get
\begin{equation}
\label{R2W}
\int d^4x
\,\,C_{{h}}(\Lambda,{\phi})\,({R^+})^2  ({\mathcal{F}^{+}})^{2h-2}~.
\end{equation}
To compute {$c_{k,h}$}, one can use constant values
${\Phi}\to {a}$ and ${W^+_{\mu\nu}}\to {f_{\mu\nu}}$; we will use the values
in Eq. \ref{floc} that correspond to the {localization deformation}.

As we remarked above, $Z^{(k)}({a},{\varepsilon})$ does not depend on
{$\bar\varepsilon$}.
However, ${\bar\varepsilon}=0$ is a limiting case: some care is needed. 
In fact, while $\mathcal{F}^{(k)}({a};{\varepsilon})$ is well-defined, 
$S^{(k)}[{a};{\varepsilon}]$ diverges because of the (super)volume integral 
$\int d^4x \,d^4\theta$. The presence of
{$\bar\varepsilon$} regularizes the superspace integration by a Gaussian term,
leading to the effective rule: 
$
\int d^4x\, d^4\theta \to 1/\varepsilon^2
$;
one can then work with the {effective action}, i.e., the \emph{full} 
{instanton partition function}. Moreover,
{$a$} and {$\varepsilon,\bar\varepsilon$} deformations localize completely the
integration over moduli space which {can thus be carried out}
\cite{Nekrasov:2002qd,Flume:2002az,Losev:2003py,Nekrasov:2005wp}.

With {$\bar\varepsilon\not=0$} (complete localization) a {trivial} superposition of 
instantons of charges $k_i$ contributes to the sector $k= \sum k_i$;
such {disconnected} configurations do \emph{not} contribute 
when {$\bar\varepsilon=0$}. 
The partition function computed by localization corresponds thus to the 
{exponential} of the non-perturbative prepotential:
\begin{equation*}
\label{ZvsF}
Z({a};{\varepsilon})  =  
\exp\left(\frac{\mathcal{F}_{\mbox{\tiny n.p.}}({a},{\varepsilon})}{{\varepsilon^2}}\right)
= \exp\left(\sum_{k=1}^\infty \frac{\mathcal{F}^{(k)}({a},{\varepsilon})}{{\varepsilon^2}}\right)
=  \exp\left(\sum_{h=0}^\infty \sum_{k=1}^\infty {c_{k,h}} \frac{{\varepsilon^{2h-2}}}{{a^{2h}}}
\left(\frac{\Lambda}{{a}}\right)^{4k}
\right)~.
\end{equation*}
In conclusion, the computation via localization techniques of the multi-instanton partition function
 $Z({a};{\varepsilon})$ determines the coefficients ${c_{k,h}}$
 which appear in the {gravitational} $F$-terms of the $\mathcal{N}=2$ effective action
\eq{R2W}
via the expression of $C_{{h}}(\Lambda,{\phi})$ given in \eq{Fnp}.

The very same {gravitational} $F$-terms can been extracted in a completely 
different way by considering
{topological string} amplitudes on suitable {Calabi-Yau} manifolds. 
Indeed, the construction that goes under the name of 
{geometrical engineering} embeds directly the SW low energy description of
$\mathcal{N}=2$ SYM theory into {string theory} as {type IIB} on a suitable {local CY} manifold {$\mathfrak{M}$}
\cite{Kachru:1995fv,Katz:1996fh}. In this realization, 
the geometric moduli of {$\mathfrak{M}$} encode the gauge 
theory data ($\Lambda, {a}$), and the coefficients $C_{{h}}$ in the 
l.e.e.a. gravitational F-terms
are given by {topological string amplitudes} at genus ${h}$
\cite{Bershadsky:1993cx,Antoniadis:1993ze}. 

The two different roads to determine the $F$-couplings of \eq{R2W} must lead to the same result.
This is a very natural way to state the conjecture by N. Nekrasov \cite{Nekrasov:2002qd}
that the coefficients arising in the $\varepsilon$-expansion of multi-instanton partition
functions match those appearing in higher genus topological string amplitudes
on CY manifolds.

\begin{acknowledgement}
I thank the organizers of the Workshop ``ForcesUniverse 2006'' and
the co-authors of \cite{nuovo}. Work partially supported by the E.C. Human Potential
Programme, contract MRTN-CT-2004-005104, and by the Italian MIUR, contract PRIN-2005023102. 
\end{acknowledgement}

\end{document}